\documentclass[review]{elsarticle}
\usepackage[export]{adjustbox}
\usepackage{lineno,hyperref}
\usepackage{color}
\usepackage{amssymb,amsmath}

\journal{Journal of \LaTeX\ Templates}

\begin{document}

\begin{frontmatter}

\title{The antinucleus annihilation reconstruction algorithm of the GAPS experiment. }

\author{R. Munini$^{a}$ $^{b}$, E. Vannuccini$^{c}$,  M. Boezio$^{a}$ $^{b}$ , P. von Doetinchem$^{d}$, C.~Gerrity$^{d}$, A. Lenni$^{e}$ $^{a}$ $^{b}$, N. Marcelli$^{f}$ $^{g}$, S. Quinn$^{h}$, F. Rogers$^{i}$, J.L. Ryan$^{h}$ ,  A. Stoessl$^{d}$, M. Xiao$^{i}$, N. Saffold$^{j}$, A. Tiberio$^{c}$, M.~Yamatani$^{k}$  }

\address{$^{a}$INFN, Sezione di Trieste, I-34149 Trieste, Italy}
\address{$^{b}$IFPU, I-34014 Trieste, Italy}
\address{$^{c}$INFN, Sezione di Firenze, I-50019 Sesto Fiorentino, Florence, Italy} 
\address{$^{d}$Department of Physics and Astronomy, University of Hawaii at Manoa, 2505 Correa Rd, Honolulu, HI 96822, USA
}
\address{$^{e}$Universita` di Trieste, I-34127, Trieste, Italy}
\address{$^{f}$University of Rome ``Tor Vergata'', Department of Physics, I-00133 Rome, Italy}
\address{$^{g}$INFN, Sezione di Rome ``Tor Vergata'', I-00133 Rome, Italy} 
\address{$^{h}$Department of Physics and Astronomy, University of California at Los Angeles, Los Angeles, CA 90095, USA}
\address{$^{i}$Department of Physics, Massachusetts Institute of Technology, 77 Massachusetts Ave, Cambridge, MA 02139, USA}
\address{$^{j}$Columbia Astrophysics Laboratory, Columbia University, 550 W 120th St, New York, NY 10027, USA}
\address{$^{k}$Institute of Space and Astronautical Science, Japan Aerospace Exploration Agency (ISAS/JAXA), Sagamihara, Kanagawa 252-5210, Japan} 

\begin{abstract}
The General AntiParticle Spectrometer (GAPS) is an Antarctic balloon-borne detector designed
to measure low-energy cosmic antinuclei ($< 0.25$\,GeV/$n$), with a specific focus on antideuterons, as a distinctive signal from dark matter annihilation or decay in the Galactic halo. 
The instrument consists of a tracker, made up of ten planes of lithium-drifted Silicon Si(Li) detectors,
surrounded by a plastic scintillator Time-of-Flight system.
GAPS uses a novel particle identification method based on exotic atom capture and decay with
the emission of pions, protons, and atomic X-rays from a common annihilation vertex. 

An important ingredient for the antinuclei identification is the reconstruction of the "annihilation star" topology. 
A custom antinucleus annihilation reconstruction algorithm, called the "star-finding" algorithm, was developed to reconstruct the annihilation star fully,  
determining the annihilation vertex position and reconstructing the tracks of the primary and secondary charged particles. 
The reconstruction algorithm and its performances were studied on simulated data obtained with the {\tt Geant4}-based GAPS simulation software, which fully reproduced the
detector geometry. 
This custom algorithm was found to have better performance in the vertex resolution and reconstruction efficiency compared with a standard Hough-3D algorithm.
\end{abstract}


\end{frontmatter}

\section{Introduction}

Cosmic-ray antinuclei offer a unique opportunity to  
probe a variety of dark matter models that evade collider, direct, or other indirect searches (\cite{Doetinchem_2020} and references therein for a recent overview).

Since their first detection in the 1970s, antinuclei, namely antiprotons, have been used to investigate dark matter models and constrain cosmic-ray production and propagation models. While the results on cosmic-ray antiprotons are mostly consistent with secondary production, a possible contribution from dark matter annihilation or decay cannot be excluded \cite{Cuo17, Cui17}. 

In the early 2000s \cite{Donato2000}, it was realized that heavier antinuclei, in particular antideuterons, could be a significantly cleaner signature of dark matter. 
In fact, due to the kinematics of the antinuclei formation, at low energies, the antideuteron astrophysical background is expected to be much lower than that of antiprotons, and a variety
of dark matter models predict an antideuteron flux exceeding the background by orders of magnitude in the energy range below a few\,GeV/$n$ (e.g., \cite{For13}).
Cosmic antideuterons have not yet been observed, with only upper bounds being published (e.g., \cite{BESS_AntiD}). 
Any detection would open a new field of cosmic research.

Recently, the AMS-02 collaboration announced several
candidate events with mass and charge consistent with antihelium nuclei~\cite{AMS02_AntiHe}. Data taking, analyses, and interpretation of these events are still ongoing. If confirmed, these results would be an exciting sign of new physics challenging the present knowledge of particle physics and cosmology.

All existing antinuclei results have been obtained with instruments designed for antiparticle identification based on a magnetic spectrometer. 
The General AntiParticle Spectrometer (GAPS) \cite{Mori2002,Hailey2009} will be complementary to these apparatuses. 
GAPS is the first experiment specifically optimized for low-energy $(<0.25$\,GeV/$n$) cosmic antinuclei detection.  
GAPS has adopted a novel antinuclei detection technique based on slowing down incoming antinuclei and observing the subsequent exotic atom formation, decay, and annihilation signature.  

The GAPS program envisions at least three Antarctic balloon flights, the first one scheduled for the austral summer of 2022-2023. 
The flight-path location, at low geomagnetic cutoff, is ideal for the study of low-energy cosmic particles. 
GAPS will collect the largest statistics of low-energy antiprotons to date, extending the existing measurements to unexplored low energies ($< 100$ MeV), and will improve sensitivity to heavier antinuclei by at least two orders of magnitude. Detailed discussion of the antinuclei sensitivity can be found in \cite{Ara16,Saffold2021}.
This paper describes the algorithm developed to reconstruct the annihilation topology that is an important ingredient for the antinuclei identification, e.g. see \cite{Ara16}.

\section{The GAPS Detector}
\label{Sec1}

\begin{figure}[th]
\centering
\includegraphics[width=1\textwidth]{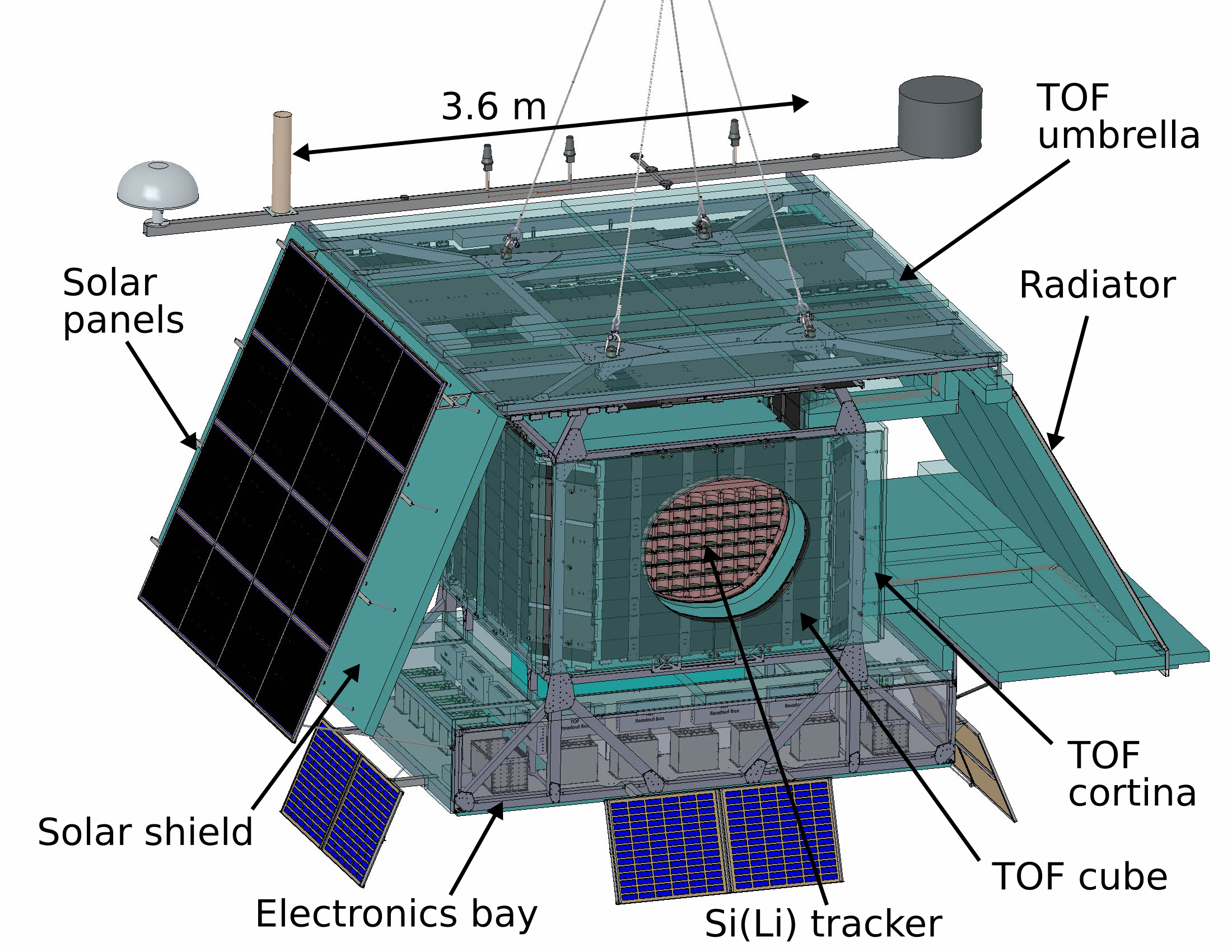}
\caption{Schematic view of the GAPS detector. }
\label{fig1}
\end{figure}

Figure ~\ref{fig1} shows a schematic view of the GAPS apparatus consisting of a Time-of-Flight (ToF) and tracking systems. 

The ToF system consists by $\sim 160$ plastic scintillator  paddles \cite{ToF_proceeding} and is arranged in and outer ToF system and an inner ToF system (see Figure ~\ref{fig1}).  The
  outer ToF consists of an umbrella of scintillator oriented horizontally and
  a cortina of four walls of scintillator oriented vertically. The inner ToF is
 a cube of scintillator, consisting of four sides, a top and a bottom.
Each plastic scintillator paddle is 6.35 mm thick and 16 cm wide. The umbrella consists of 1.8 m and 1.51 m length paddles, whereas the cortina and the cube use 1.72m and 1.51, 1.8 m and 1.56 m lengths respectively. 
The ToF system measures the time information necessary to reconstruct velocity
of particles and the ionization energy losses d$E$/d$x$ of particles.
The ToF also provides the overall trigger for GAPS.

The core of the apparatus is a tracking system  made of ten planes of cylindrical Si(Li) detectors
\cite{Perez2018,kozai,Rogers_2019,Saffold2021b}. On each supporting plane, made of aluminum, the Si(Li) cylinders are arranged in a $6 \times 6$ 
array of modules, each with four Si(Li) detectors read-out by a dedicated ASIC \cite{Asic_proceeding}. 
Each Si(Li) detector, with overall dimensions $\sim 10 $ cm-diameter and $\sim 2.5 $ mm-thick, is segmented into eight strips of equal area. 
An oscillating heat pipe system \cite{Fuke2017,Okazaki2018} 
is used to cool the Si(Li) detectors to the requisite operational temperature, $\approx -40 ^{\circ}$ C.
A precise reconstruction of the annihilation topology is one of the key goals of the tracker. 

\begin{figure}[h]
\centering
\includegraphics[width=.71\textwidth]{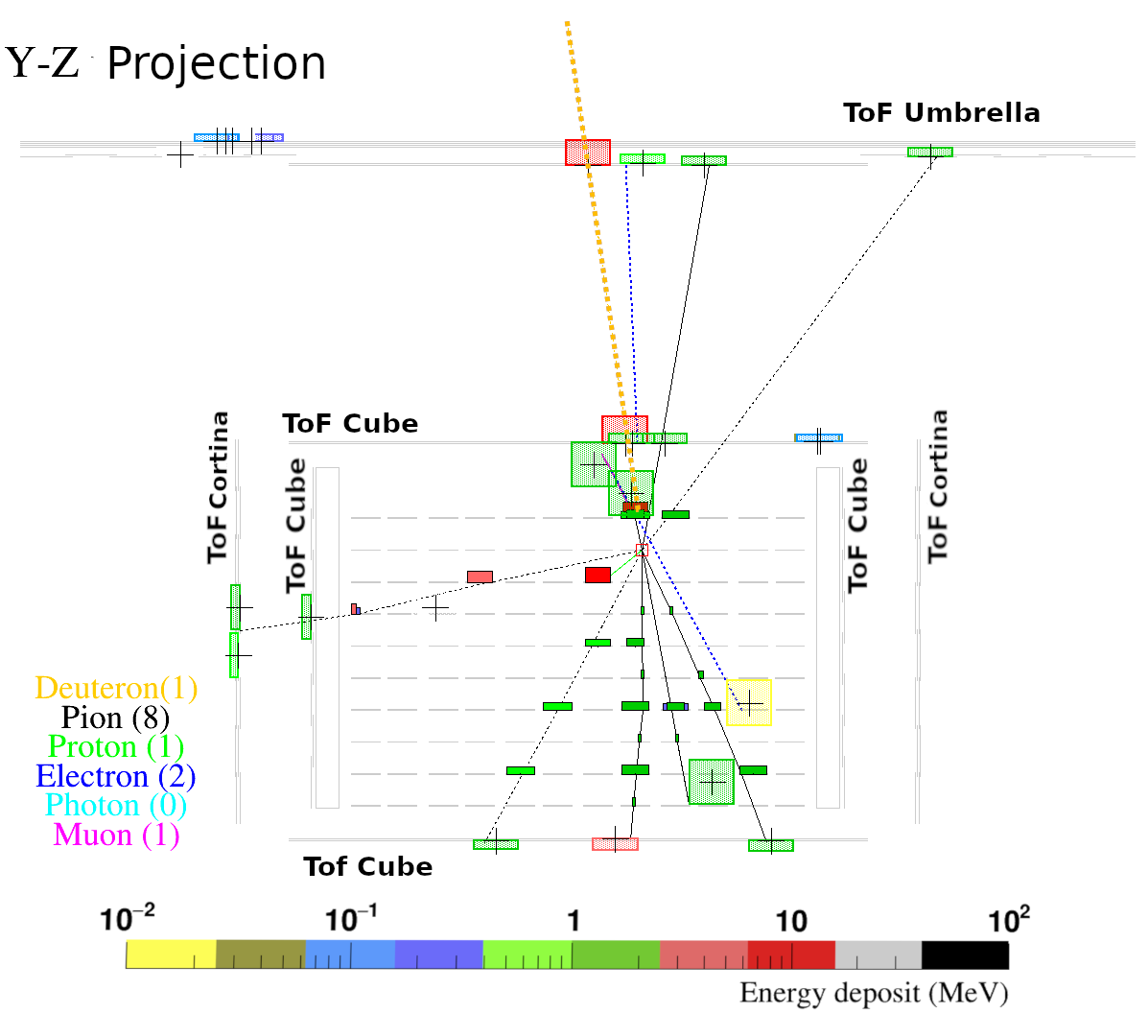}
\caption{
A simulated antideuteron with $\beta = 0.28$. 
The panel represents a two dimensional projection (from the side) of the detector. The lines represent the Monte Carlo trajectories (solid lines particles, dotted lines antiparticles) and each color is associated with a different particle species (legend on the left bottom corner). The energy released by a particle in an active volume is represented by the colored boxes (color code in the palette at the bottom of the figure).
The crosses inside the ToF hit boxes represent the one sigma spatial resolution and are centered on the estimated position of the hit.
The height of the ToF hit boxes seen edge-on is proportional to the deposited energy.
The width of the tracker hit boxes represents the dimension of the strip, and the height of the box is proportional to the deposited energy. The open red square indicates the Monte Carlo annihilation position.}
\label{fig1a}
\end{figure}

The incoming antinuclei, hereafter called primary particles, that have been sufficiently slowed by ionization losses can form an exotic atom by replacing a shell electron in a silicon or aluminum atom. 
Subsequently, the exotic atom decays with a series of atomic transitions emitting X-rays of specific energies~\cite{Ara16}. Finally, the primary particle annihilates with the target nucleus producing secondary particles, primarily pions and protons,
from a common vertex. 
The typical topology, without the atomic X-ray contribution, from an antideuteron annihilation is shown in Figure ~\ref{fig1a} (see the caption for a detailed description). The lines show the Monte Carlo trajectories of the particles (the primary antideuteron is the dotted yellow line), while the boxes represent the hits, i.e., the digitized energy depositions in each active volume. This work describes the algorithm developed to reconstruct this topology. 

Considering the various particle and antiparticle Galactic cosmic-ray abundances, e.g., \cite{Boezio2020}, a rejection power of at least of $10^{6}$ is required to separate low-energy antiprotons from the much more abundant cosmic-ray particle background.
To extract a possible antideuteron signal from the standard antiproton component,
an additional $\sim 10^{5}$ rejection factor is necessary.
Antinuclei indentification 
will be performed combining 
several quantities related to the kinematics of the primary particle, such as the range and ionization losses, the energy of the X-rays, and the number of secondary particles resulting from the annihilation \cite{Ara16}.  
Critical is the identification of the annihilation as well as the precise determination of the range and ionization losses of the primary particle. 
For example,  
the range in the GAPS tracking system of antideuterons and antiprotons with nearly vertical incidence and comparable velocity $\beta < 0.4$ ($\beta = v /c $, with $v$ the particle velocity and $c$ the speed of light) differ on average by more than 12 cm. 
Thus, it is essential to reconstruct 
the annihilation topology
with high precision.

\section{Instrument Simulation}
\label{SimInstr}
The reconstruction algorithm was developed using the  GAPS simulation software, which reproduces the instrument geometry and materials using the {\tt Geant4} toolkit \cite{Geant4}. A set of around $10^5$ simulated protons, antiprotons, antideuterons, and antihelium-3 nuclei were used to develop the reconstruction. 

A realistic instrument response for time, energy and position measurements was introduced.
In this work, the position resolution along the ToF paddle length was assumed at $\sigma_d = 4 $\,cm,  based on the end-to-end ToF paddle timing  resolution $\sigma_{t}$ (ToF paddles are read out by silicon photomultipliers placed at both ends of the paddle) measured to be better than $400$\,ps, while the paddle width (16 cm) and thickness (0.635 cm) defined the precision in the two other coordinates.
The mechanical positions and the dimensions ($\approx 1.25$\,cm $\times$ $\approx 10$\,cm $\times$ $0.25$\,cm) of the Si(Li) strips determined the spatial information and resolution for the positions in the tracking system.

 The tracker energy resolution is dominated by the ADC resolution above a few MeV. Therefore, the energy resolution was implemented by converting the true energy deposition to ADC digits and back.
The transfer functions of each channel on an ASIC board were experimentally measured. The simulated energy was converted through the transfer function to ADC digits. The value obtained was rounded to the nearest integer and converted back to energy with the same function. The final total tracker energy resolution (Full Width Half Maximum) is about 4 keV (largely due to the Gaussian electronic noise) \cite{Asic_proceeding}  up to a few hundred keV of energy deposition, slowly increasing up to about $100$ keV at energy depositions higher than $50$\,MeV.

Since adjacent ToF paddles overlap by approximatively one centimeter, 
about $20\%$ of the trajectories have multiple hits
(up to four) in the outer ToF and cube combined. 
For this reason, a general clustering procedure was implemented. 
Two ToF hits were clustered if their distance was consistent within $3 \sigma_{d}$ and their absolute time difference was consistent within $3 \sigma_{t}$. In this way, the possibility to wrongly associate hits due to a secondary particle are drastically reduced. The time and the position of the resulting hit is the average time and position of the clustered hits and its energy is the arithmetic sum of the energy depositions. The associated position error is reduced accordingly. 

Two trigger configurations were implemented in the simulation: a minimum bias trigger, requiring at least one hit in the outer ToF and at least one hit in the inner ToF, and a second trigger configuration requiring a total of at least eight hits in the ToF system with at least three hits in the outer ToF and three hits in the inner ToF. 
In addition, this last trigger scheme requires that the TOF energy depositions are in the range of slow-moving $|$Z$|=1$ or $|$Z$|=2$ to reject minimally ionizing (high-velocity) and high-charge particles.
This second configuration was optimal for selecting  the annihilating antinuclei with an efficiency higher than $60\%$ in 
the $\beta$ range of scientific interest for the GAPS experiment,
$\sim 0.2$ to $\sim 0.6$,
while reducing the trigger rate down to less than 1\,kHz from the expected MHz rate with the minimum bias trigger \cite{ToF_proceeding}.

\section{Antinucleus Annihilation Reconstruction}

The antinucleus annihilation reconstruction algorithm was designed to determine the annihilation vertex of antinuclei that annihilate in the tracker volume and to reconstruct the primary track as well as the secondary tracks originating from the  
vertex. 

First, the primary track was identified, followed by the secondary tracks associated with the primary track. Finally, primary and secondary tracks were used to estimate the position of the annihilation vertex.  

\subsection{Primary Track Finding} 

The primary track was identified using the timing information provided by the ToF system.
It was assumed that the primary was the first particle to hit the outer ToF and the inner ToF. Consequently, the hits with the shortest time distance from the trigger signal in the outer and the inner ToF were selected as hits on the primary track.
Then, a track-following approach was adopted. 
The two hits were used to construct an initial track segment, which was extrapolated into the silicon tracker detection planes to include further hits progressively, as discussed below.

A small fraction of primary events (at the level of a few per thousand) produced delta rays in the outer ToF. Since delta rays are  relativistic, i.e. with a  $\beta \approx 1$, they almost always reached the cube ahead of the 
primary. An energy threshold was imposed on the selected cube hit to reduce this contamination to a negligible amount. 
From Monte Carlo, it was estimated that more than $99.9\%$ of the delta rays produced by the primary particle released less than $1$\,MeV inside the cube. In contrast, $99.9\%$ of the primary antinuclei deposited $>1$\,MeV. For this reason 
the cube point associated with the primary particle had to satisfy: $E_{\text{cube}} > 1$\,MeV. 
 
The tracker hits were required to satisfy the following three criteria to be associated with the primary track:

\begin{enumerate}

  \item {\bf Spatial consistency:}
  From the  initial guess of the primary trajectory, obtained by linearly interpolating the ToF hits in the outer ToF and the cube, 
  the impact position $\mathbf{p}_{1}$ 
  on the first tracker plane crossed by the projected track (this plane will be referred as plane 1 hereafter)  
  was calculated along with
  its associated uncertainty $\delta_{1}$. This is the uncertainty that resulted from the propagation of the 
  position errors of the interpolated hits. Moreover, 
  the projected linear displacement $\delta_{\text{ms}}$ with respect to $\mathbf{p}_{1}$
  due to the multiple scattering process was calculated (based on the reconstructed $\beta$ and assuming proton mass) and linearly added to $\delta_{1}$ to obtain the $\delta_{\text{tot}}$.
  This approach for the scanning region was found to be the best balance between a high-efficiency inclusion of primary hits and a reliable rejection of the spurious hits.
  Then, the distances $d_i$
  between each of the hits on this first tracker plane crossed by the projected track and $\mathbf{p}_{1}$ were calculated. 
  If $d_i < \alpha_1(\beta) \cdot \delta_{\text{tot}}$ the hit was selected, with $\alpha_1(\beta)$ being a distance-related, $\beta$-dependent coefficient function for plane 1 that will be discussed together with the other cuts later on in 
  this section .
  \item {\bf Energy consistency:} The energy depositions of the various hits selected by the spatial consistency requirement in plane 1 were converted to the energy released per unit of mass thickness d$E/$d$x$, with d$x = $d$l \cdot \rho$ and d$l$ the path length, evaluated using the trajectory information, inside an active volume of density $\rho$. 
  If the selected hits belonged to adjacent strips and their energy depositions were consistent within $50\%$ they were clustered, resulting in a hit with the average position and the arithmetic sum of the clustered strips.
  Then, the hit with the highest d$E/$d$x$ was selected if this energy was consistent with the one expected from a slowing-down particle. Consequently, the selected hit had to satisfy the condition $E_{1}  > E_{0}/\nu_1(\beta)$ where  $E_1$ is the energy of the selected hit on the plane 1 and $E_{0}$ is the energy of the hit associated with the primary track in the last crossed plane, in this case the cube.
  Similar to $\alpha_1(\beta)$ above, $\nu_1(\beta)$ is a $\beta$-dependent coefficient function for plane 1, but focused on the energy deposition progression.
    
  \item {\bf Upper limit to the distance between consecutive hits:} Finally, a check on the spatial location of the hits was performed.     
  It is possible that a hit from a secondary particle is spatially and energetically consistent with the primary track even if it occurred after the annihilation vertex. In this case, the distance between the previous primary hit and the secondary hit should generally be larger than the mean distance of two consecutive actual primary hits. Therefore, a cut function 
  was introduced. The identified hit on plane 1 was associated to the primary track if its spatial distance from the nearest associated primary hit was less than $\gamma_1(\beta)$.

\end{enumerate}
 
If a hit on plane 1 satisfied all these three requirements, the hit was associated with the primary track, and a straight-line, least-squares minimization was performed on the associated primary hits (\ref{AppendixA}). The resulting trajectory was used to estimate the impact position on the following tracker plane. Hits in the second tracker plane were then associated using the corresponding $\alpha_2(\beta)$, $\nu_2(\beta)$ and $\gamma_2(\beta)$ cut functions, followed by a new straight-line fit. This procedure is repeated for all tracker layers.

The position and energy loss of the hits classified with the track finding procedure described above were compared with the position and energy loss of the hits due to the primary particle according to the Monte Carlo truth. From this comparison, the coefficients $\alpha_j(\beta)$, $\nu_j(\beta)$ and $\gamma_j(\beta)$ for each traversed tracker plane $j$ were determined to provide selection cuts with a selection efficiency of the primary hits of $\sim 98\%$. Thus, a balance between the rejection of hits that did not belong to the primary and the selection efficiency of the primary hits was ensured.
  
\begin{figure}
\centering
\includegraphics[width=1\textwidth]{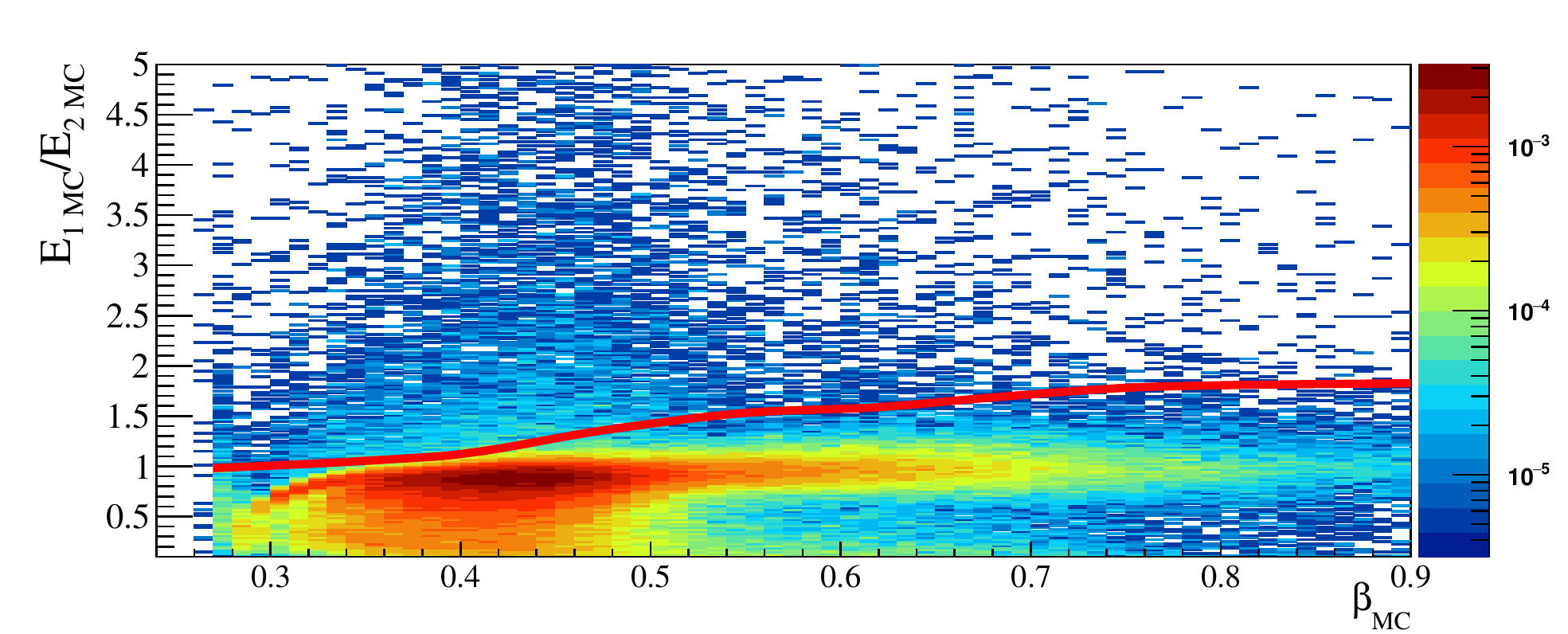}
\caption{The energy selection cut applied to the second tracker hit as a function of the Monte Carlo $\beta$. The $98\%$ efficiency cut $\nu_2(\beta)$ is represented by the solid line.}
\label{fig2}
\end{figure}

Figure ~\ref{fig2} shows the ratio of the energy deposited by the primary in plane 1 and plane-2 according to the Monte Carlo, $\frac{E^{\text{MC}}_1 }{E^{\text{MC}}_2} $, as a function of the Monte Carlo velocity of the primary particle $\beta_{\text{MC}}$ . The solid line is $\nu_2(\beta)$ chosen to have a $98\%$ selection efficiency. Because of the fluctuations in the ionization
energy losses, the values of $\nu_2$, and in general all $\nu_N$, are not strictly smaller than one. The highest values of $\frac{E^{\text{MC}}_1}{E^{\text{MC}}_2} $ in Figure ~\ref{fig2} refer to primary particles traversing only a fraction of the active volume (for example, tracks very close to the detector edges\footnote{Because of the limited angular resolution of the reconstructed trajectory, the path length $x$ was evaluated assuming that the particle traversed the entire volume.}) or to primary particles with the Bragg peak occurring prior to stopping in the active volume.   

\begin{figure}[h]
\centering
\includegraphics[width=.71\textwidth]{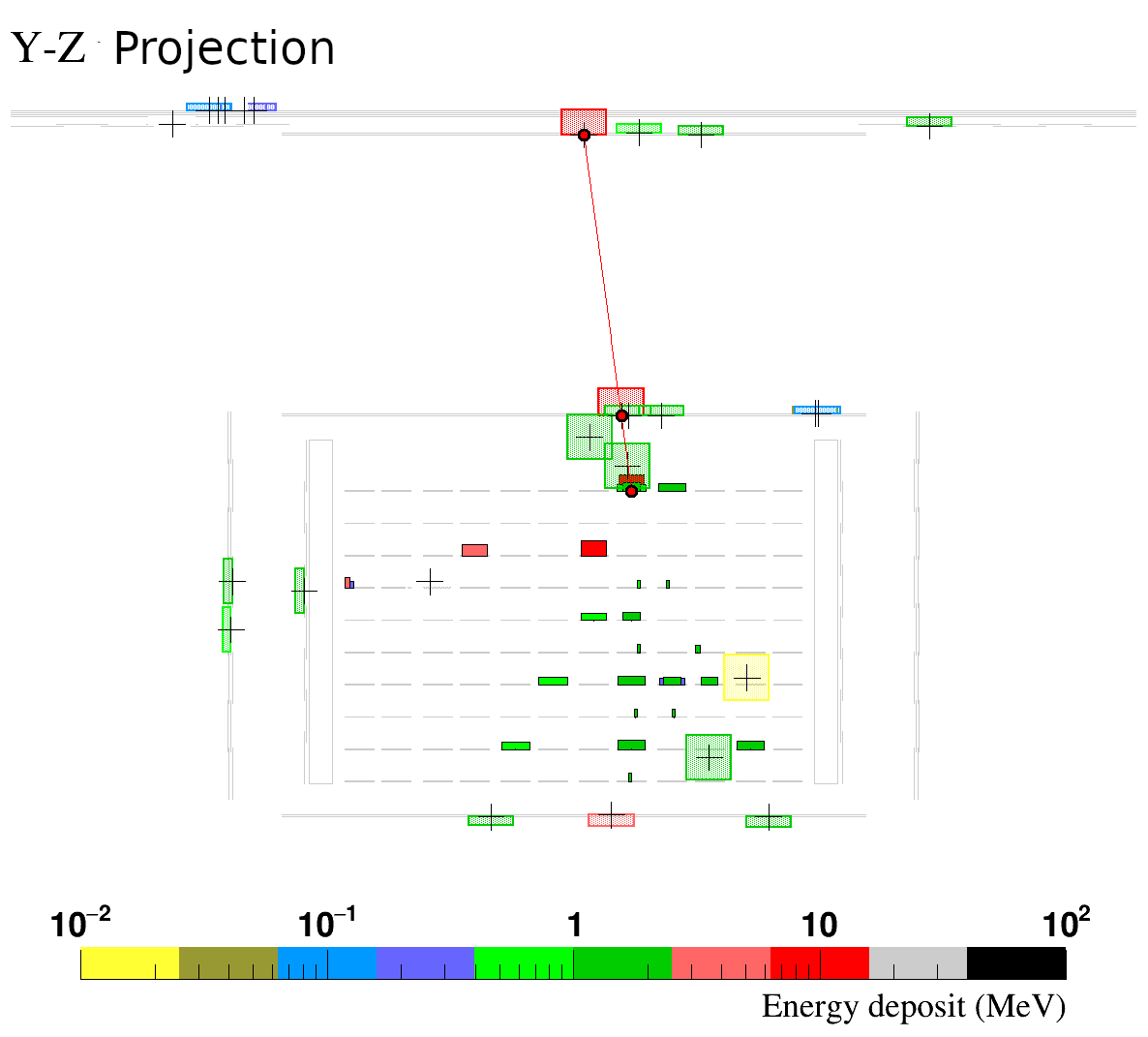}
\caption{  The same event as in Figure \ref{fig1a} with the reconstructed primary trajectory and the hits associated with the primary track (red line and circles).}
\label{fig3}
\end{figure}

Figure \ref{fig3} panel displays the same event as in Figure \ref{fig1a} with the reconstructed primary track and associated hits (red circles and solid line). 
 
If sufficiently energetic, the primary track may exit the tracker volume without annihilating, thus crossing once more the ToF system (either the cube or the cortina for very inclined track). For this reason, the search for primary hits was also extended beyond the tracker volume. In this case, the three checks were applied to the remaining ToF points. If one or more points satisfied the primary conditions, they were added to the primary track, and the least square minimization was applied again to refine the trajectory. 

\subsection{Secondary Track Finding}

After the association of hits to the primary track, secondary tracks were searched among the remaining hits. 
Several approaches were studied. This paper reports on the algorithm that was specifically developed for this experiment to search for star-like patterns. 
A global approach based on the Hough-3D transform algorithm was also developed (\ref{Hough3DSec}) and compared to the custom algorithm. The custom algorithm proved to have better efficiency and vertex resolution (Secs.~\ref{secefficiecy} and \ref{SecVertexRes}), and thus was adopted as the main reconstruction algorithm for secondary tracks. 
 
\subsubsection{ Star Finding }
\label{StarFinding}
In order to find secondary track candidates with a minimal disturbance from random hit association, this method was developed to specifically search for star-like hit patterns.

The secondary search is performed via a spatial scan along the primary direction starting from the entrance position of the primary track inside the tracker volume up to its projected exit position. The segmentation of the scan along the primary track is set to $2$\,cm. 

At each step, from the position $\mathbf{p}$ on the primary trajectory $1281$ trajectories, isotropically distributed over the solid angle, are projected, and the intercepted hits are associated.
The association interception accounted for the hit position error and the finite size of the step. 
Track candidates are obtained by iteratively selecting the trajectory with the largest number of associated hits and removing the hits up to when no trajectories intercepting at least two hits are left.
Subsequently, for the $N(\mathbf{p})$ track candidates a quantity $Q(\mathbf{p})$ is evaluated as: 
\begin{equation}
Q(\mathbf{p}) = \prod_{k=1}^{N(\mathbf{p})} \frac{n_k}{n}, 
\label{q}
\end{equation}
with $n_k$ the number of hits intercepted by each track candidate and $n$ the total number of hits of the event. Then, the minimization of $-\log{Q(\mathbf{p}})$ provides the position $\mathbf{p}_{\text{vert}}$ from which the minimum number of projected trajectories intercepted the largest number of hits. 
Finally, the secondary track candidates are obtained with 
straight-line fits to the sets of 
hits associated to each of these
projected trajectories (\ref{AppendixA}).

\subsection{Vertex Estimation}
\label{verticestimation}
The secondary track candidates obtained with the previous procedure is further analyzed to estimate their origin position, which is then identified as the annihilation vertex. The minimum requirement for the vertex reconstruction is the existence of one primary track with two hits and one secondary track with two hits.

The secondary track candidates together with the primary track are used to estimate the
annihilation vertex $\mathbf{p}_{\text{vert}}$ by minimizing the following quantity:
\begin{equation}
    \chi^2_{\text{vert}} = \sum_{k=1}^N \left[ \frac{d_k}{\delta_k} \right]^2
\end{equation}
with $N \geq 2$ number of tracks, $d_k(\mathbf{p}_{\text{vert}})$ the distance of the $k^{\text{th}}$ track 
from $ \mathbf{p}_{\text{vert}}$ and $\delta_k$ the error evaluated from the covariance matrix of the track parameters.
This minimization, along with the developed track finding procedures, significantly reduced biases in the 
vertex estimation due to spurious tracks. 
However, 
a wrong association of spurious hits could still occur. 
In order to reduce these occurrences, an iterative procedure is introduced:

\begin{itemize}

\item The point that minimized the distances from all the reconstructed tracks is chosen as the initial vertex position.
This choice is justified by the fact that the track finding algorithms constrained the track search to approximately converging tracks.

\item If the distance of a secondary track candidate from the vertex is greater than $\alpha_{\text{vert}}$, the candidate is rejected. This requirement ensured the removal of most of the tracks reconstructed from randomly aligned hits or tracks from secondary particles not originating from the annihilation vertex.  
Based on Monte Carlo information, $\alpha_{\text{vert}}$ is set to $20$\,cm, which provides $98\%$  efficiency for identifying secondary tracks originating from the annihilation vertex. 

\item With the associated track candidates that satisfied the distance check, the annihilation position is estimated via minimization described above. 

\item The association process of the secondary tracks is repeated with the new vertex position, and a new vertex is estimated. It is found that it is sufficient to repeat this process twice to obtain convergence.  

\item If the original reconstructed primary track extended beyond the estimated vertex position, the hits associated with the primary track is reevaluated accordingly, and the primary and secondary track and vertex fittings is repeated. 

\item If the vertex position is found in an active volume, the Si(Li) strip with the largest signal is associated both to the primary and secondary tracks and 
track and vertex fittings is repeated.

\end{itemize}

\begin{figure}
\centering
\includegraphics[width=.5\textwidth]{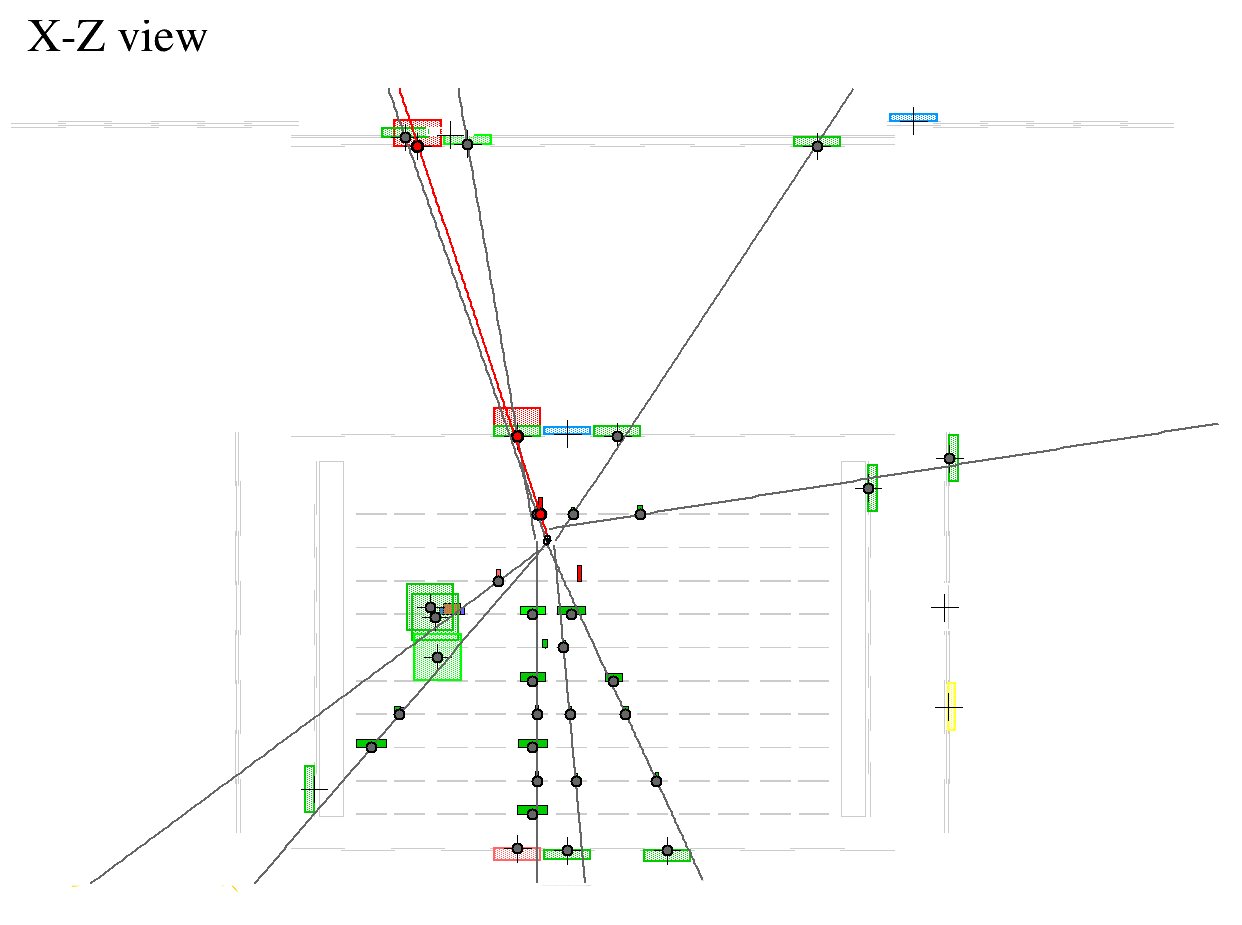}
\includegraphics[width=.5\textwidth]{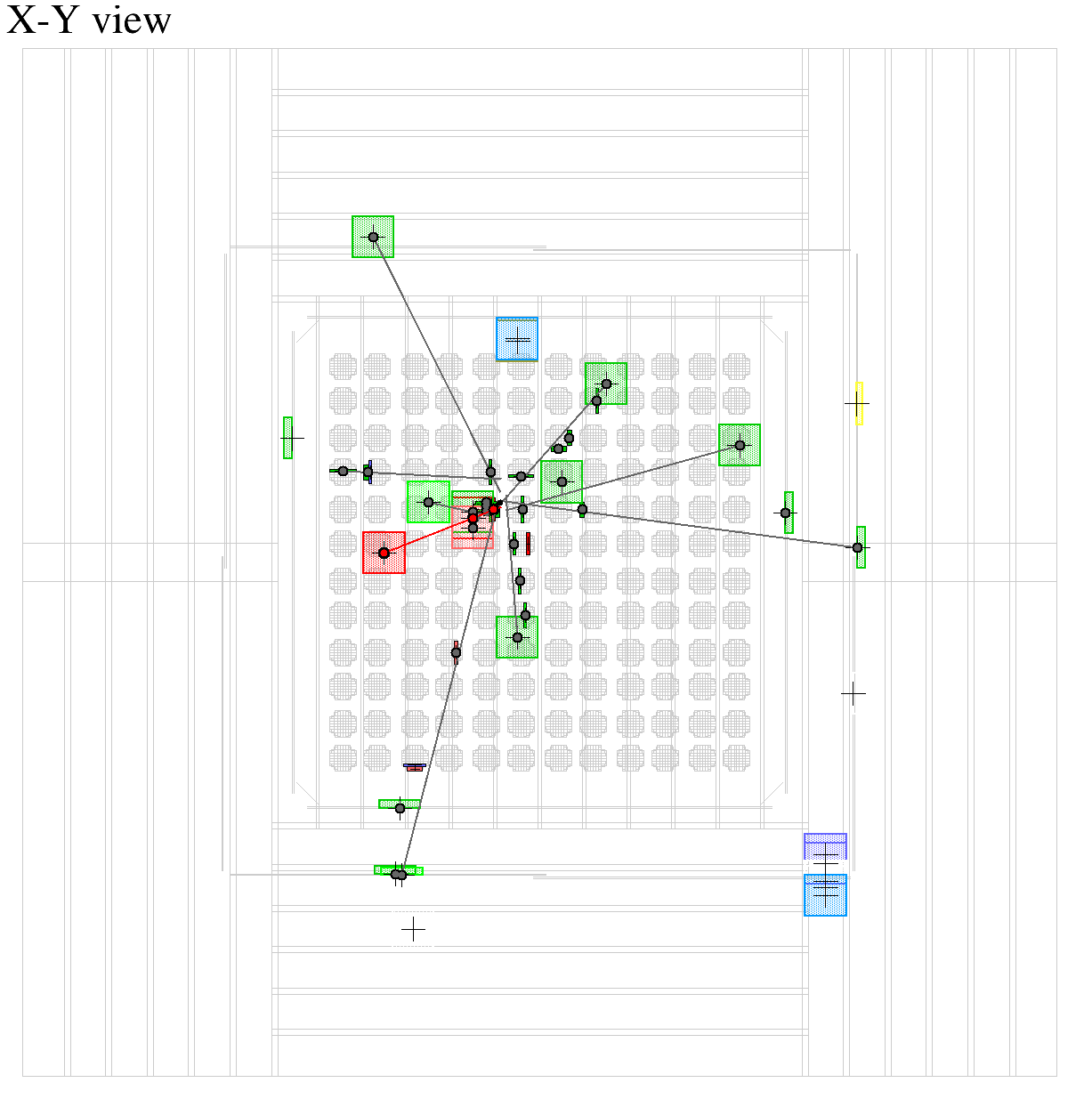}
\includegraphics[width=.5\textwidth]{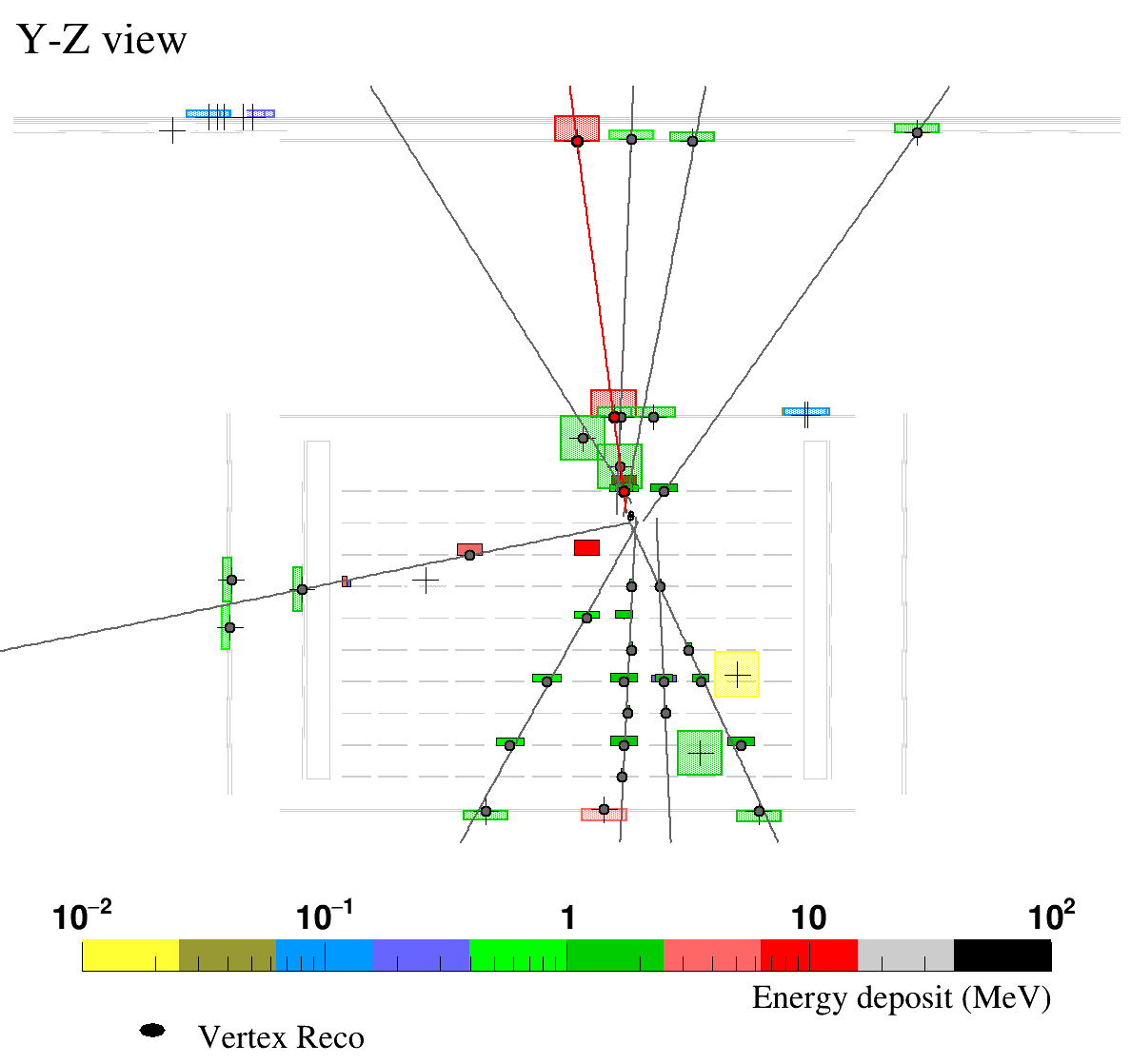}

\caption{Reconstructed antideuteron (same event displayed in Figure ~\ref{fig3}): Each panel shows a two-dimensional projection of the detector. The reconstructed secondaries found with the star-finding algorithm are represented by the gray dots and lines (nine secondaries were found). The reconstructed vertex position from the minimization procedure is represented by a black ellipse accounting for the position uncertainty (dimensions are too small to be visible on this scale). The distance between the reconstructed and the Monte Carlo vertex positions is 20\,mm.   }
\label{fig77}
\end{figure}

The minimization process provides a $\chi^2_{\text{vert}}$ and associated uncertainty. The uncertainty is defined as the $95\%$ confidence-level region of space around the reconstructed vertex. This uncertainty can be visualized as a three-dimensional ellipse centered in the estimated vertex position.
Figure ~\ref{fig77} shows the antideuteron from Figure ~\ref{fig3} with the reconstructed primary and secondary tracks (gray dots and lines) and vertex position (black ellipse). 

As described in \ref{AppendixA}, a velocity was determined for each secondary track with at least two time measurements. 
It should be noted that the time information is only provide by the ToF system. 
 For each secondary track a vertex time was estimated, by extrapolating the time along the trajectory back to the vertex position.
Then, the stopping vertex time is 
obtained by a weighted average of these estimations.
Using the vertex time, the velocities of secondary tracks with only one time measurements were also reconstructed.

\section{Reconstruction Performance}

The performance of the reconstruction algorithm was evaluated in terms of precision in reconstructing the primary track, of efficiency in identifying the annihilation of antinuclei in the tracker volume, 
and of precision in the determination of the vertex of annihilation.  
The experimental antinuclei identification capability is discussed elsewhere~\cite{Ara16}.

\subsection{Primary Reconstruction Performance}

\begin{figure}
\centering
\includegraphics[width=1\textwidth]{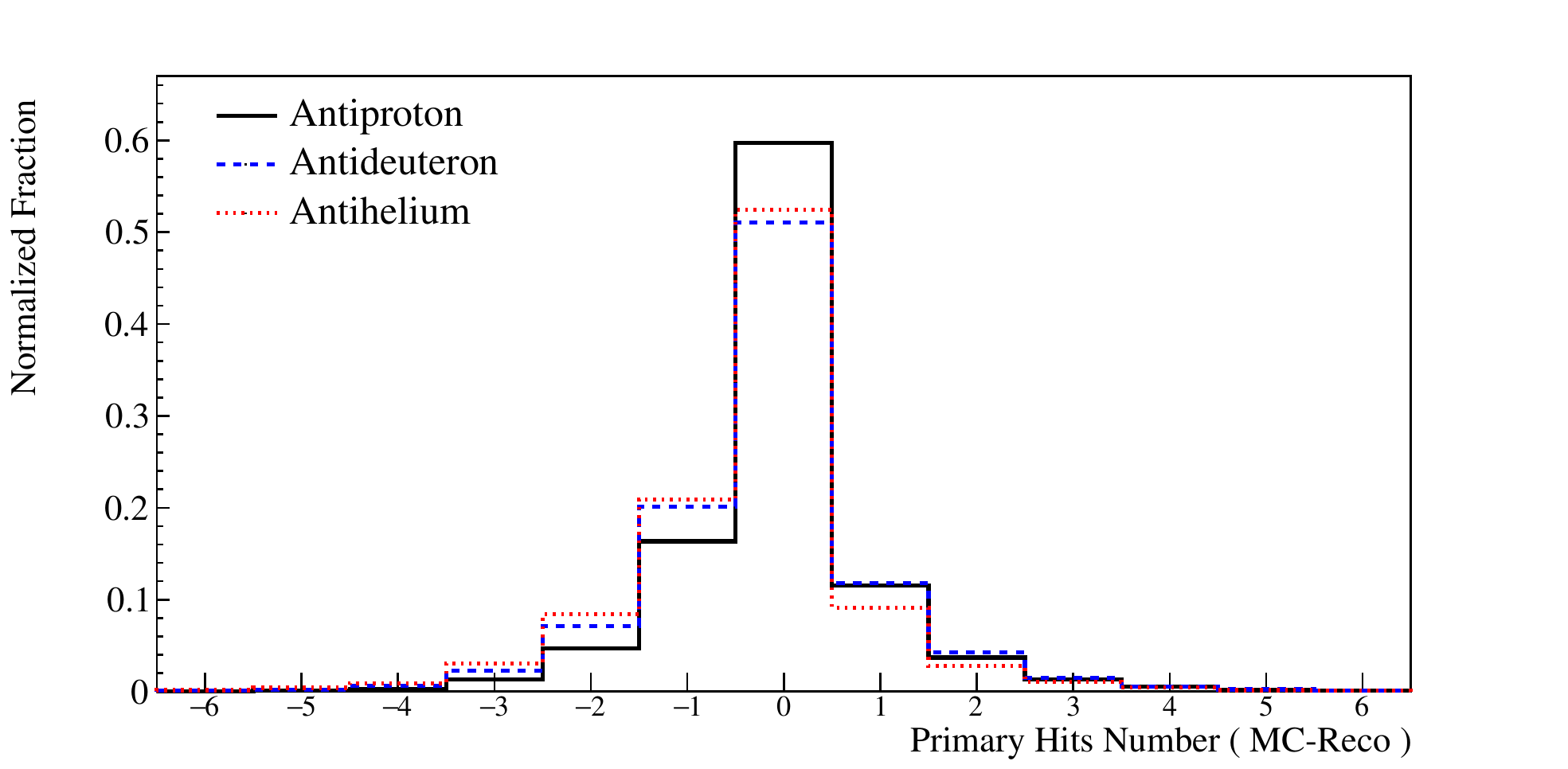}
\caption{Difference between the number of total (ToF and tracker combined) primary hits according to the Monte Carlo information and to the reconstruction for antiprotons (solid), antideuteron (long-dashed) and antihelium-3 (short-dashed).}
\label{fig5}
\end{figure}

Figure ~\ref{fig5} shows the difference between the number of primary hits obtained from the Monte Carlo and from the reconstructed information for antiprotons (solid), antideuteron (long-dashed), and antihelium-3 (short-dashed). The events were selected from those which annihilated inside the tracker volume according to the Monte Carlo information. The distributions peak at zero, indicating that for most of the events, the reconstruction procedure correctly associated the hits to the primary tracks. About $85\%$ of the events in the distributions are reconstructed with a maximum of one hit being wrongly associated.

The distributions presented in Figure ~\ref{fig5}
are skewed toward negative values indicating that the reconstruction algorithm, on average, tends to associate more hits to the primary track. The two main reasons are: 

\begin{itemize}
   \item The annihilation took place inside an inactive volume (most of the time in the Aluminum support of the Si(Li) detector) but very close to a Si(Li) detector. A few of the secondary particles crossed the Si(Li) detector, producing hits very close to the primary trajectory satisfying the spatial and energetic consistency requirements and thus being associated with the primary track. 
    \item Hits from secondary particles can satisfy the spatial and energetic consistency requirements, and thus are associated with the primary track. Since the number of secondary tracker hits is on average higher in the case of antideuterons and antihelium-3, this effect is more pronounced for antideuterons than for antiprotons.
\end{itemize}

\subsection{Reconstruction Efficiency}
\label{secefficiecy}

\begin{figure}
\centering
\includegraphics[width=1\textwidth]{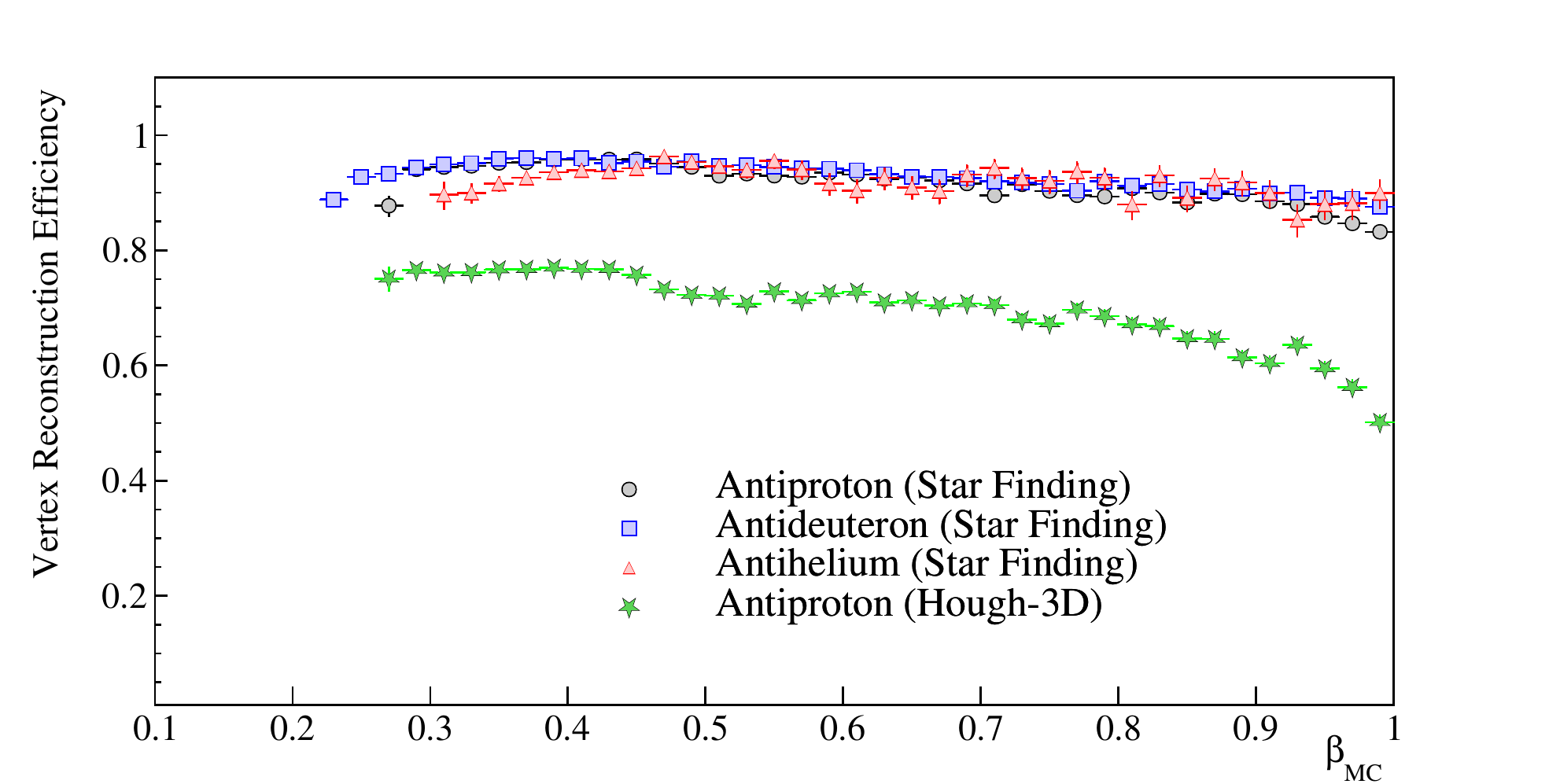}
\caption{Vertex reconstruction efficiency for the star-finding algorithm (circles antiprotons, squares antideuterons, and triangles antihelium-3). The initial sample was a set of antinuclei selected with the second trigger configuration that annihilated inside the tracker volume according to the Monte Carlo information. The green stars represent the vertex reconstruction efficiency obtained with the Hough3D algorithm.  }
\label{fig9}
\end{figure}

In Figure ~\ref{fig9} the vertex reconstruction efficiency of the star-finding algorithm is shown for different antinucleus species as a function of the generated primary particle $\beta_{\text{MC}}$. These efficiencies were calculated for a sample of antinuclei selected with the second trigger configuration (Sec.~\ref{SimInstr}) and the requirement that they annihilated inside the tracker volume according to the Monte Carlo information. The final sample of events was selected, requiring the existence of a reconstructed vertex. 
The efficiency is almost constant for all velocities at $95\%$ with a cutoff at low velocities that corresponds to the minimum energy needed for the primary to cross the ToF paddles and reach the tracker volume. For antihelium-3 nuclei, this cutoff energy is higher since the ionization energy losses are about four time higher at the same velocity. As the velocity increases, antinuclei were able to enter the tracker volume and be identified. Figure ~\ref{fig9} also shows the Hough-3D vertex reconstruction efficiency obtained with a sample of antiprotons. This efficiency is around $15\%$ lower with respect to the star-finding algorithm. 

The star-finding algorithm reconstructs the annihilation multiplicity with high efficiency. In the $\beta$ range of scientific interest of the experiment for more than $60\%$ of the annihilating antinuclei the number of reconstructed secondaries differs by not more than one from the Monte Carlo number of secondaries \footnote{The Monte Carlo number of secondaries is obtained by counting the number of pion and proton tracks originated from the Monte Carlo annihilation position and which traversed at least two active volumes (producing two hits).}.

\begin{figure}
\centering
\includegraphics[width=1\textwidth]{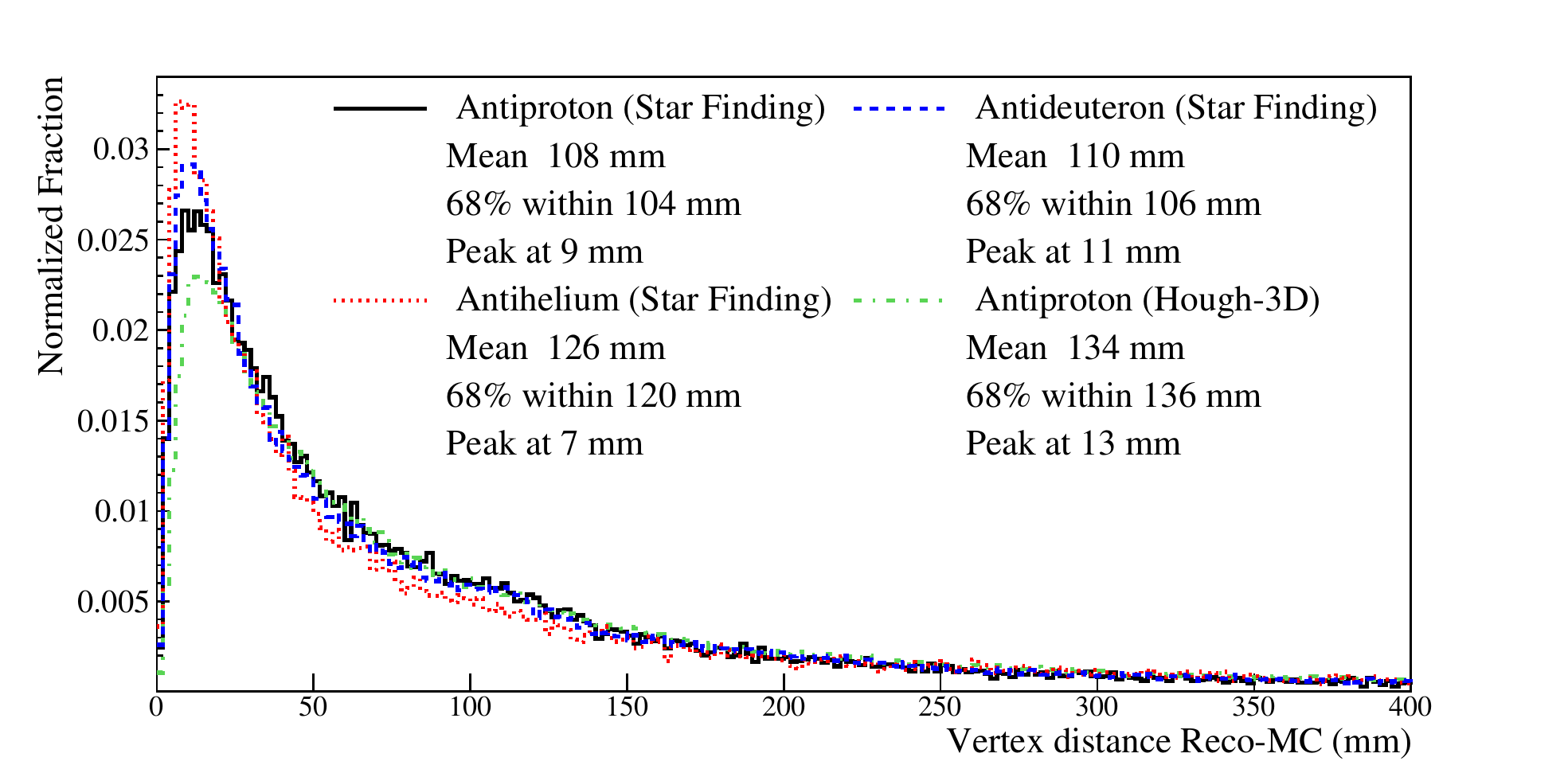}
\caption{Distribution of the absolute distance in\,mm between the Monte Carlo annihilation vertex and the reconstructed vertex. The distributions refer to simulated antiproton (solid), antideuteron (dashed), and antihelium-3 (dotted) events with a vertex inside the tracker volume according to the Monte Carlo and a reconstructed vertex with the star-finding algorithm. The distribution obtained with the Hough-3D algorithm for antiprotons (dotted-dashed) is also shown.    }
\label{fig11}
\end{figure}

\subsection{Vertex Spatial Resolution}
\label{SecVertexRes}

A good estimation of the vertex position is essential for a precise reconstruction of the trajectory and path length of the primary antinucleus. The reconstructed vertex resolution is presented in Figure ~\ref{fig11} and shows the absolute distance (in\,mm) between the Monte Carlo vertex and the reconstructed vertex positions for antiprotons (solid line), antideuterons (dashed line), and antihelium-3 (dotted line) obtained with the star-finding algorithm from a sample of events, which annihilated inside the tracker volume according to the Monte Carlo. The distributions peak at around 1\,cm and extend toward higher values. $68\%$ of the events were reconstructed with a vertex position closer to the Monte Carlo vertex than the value $d_{68\%}$ shown in the legend. 
To better appreciate these spatial resolutions, it can be noted that, on average, antideuterons with velocities $\beta < 0.4$ and entering GAPS apparatus with nearly vertical incidence annihilate at rest more than 12 cm deeper in the tracking system than antiprotons of comparable, in terms of  GAPS ToF resolution, velocities. 
Figure ~\ref{fig11}  also shows the vertex resolution for antiprotons reconstructed with the Hough-3D algorithm (dashed-dotted line). In general, with respect to the star-finding algorithm, the vertex position obtained with the Hough-3D algorithm is less accurate, for the case of antiprotons, the peak is around 13\,mm instead of 9\,mm, and $d_{68\%}$ is $136$\,mm instead of $94$\,mm.

\begin{figure}
\centering
\includegraphics[width=1\textwidth]{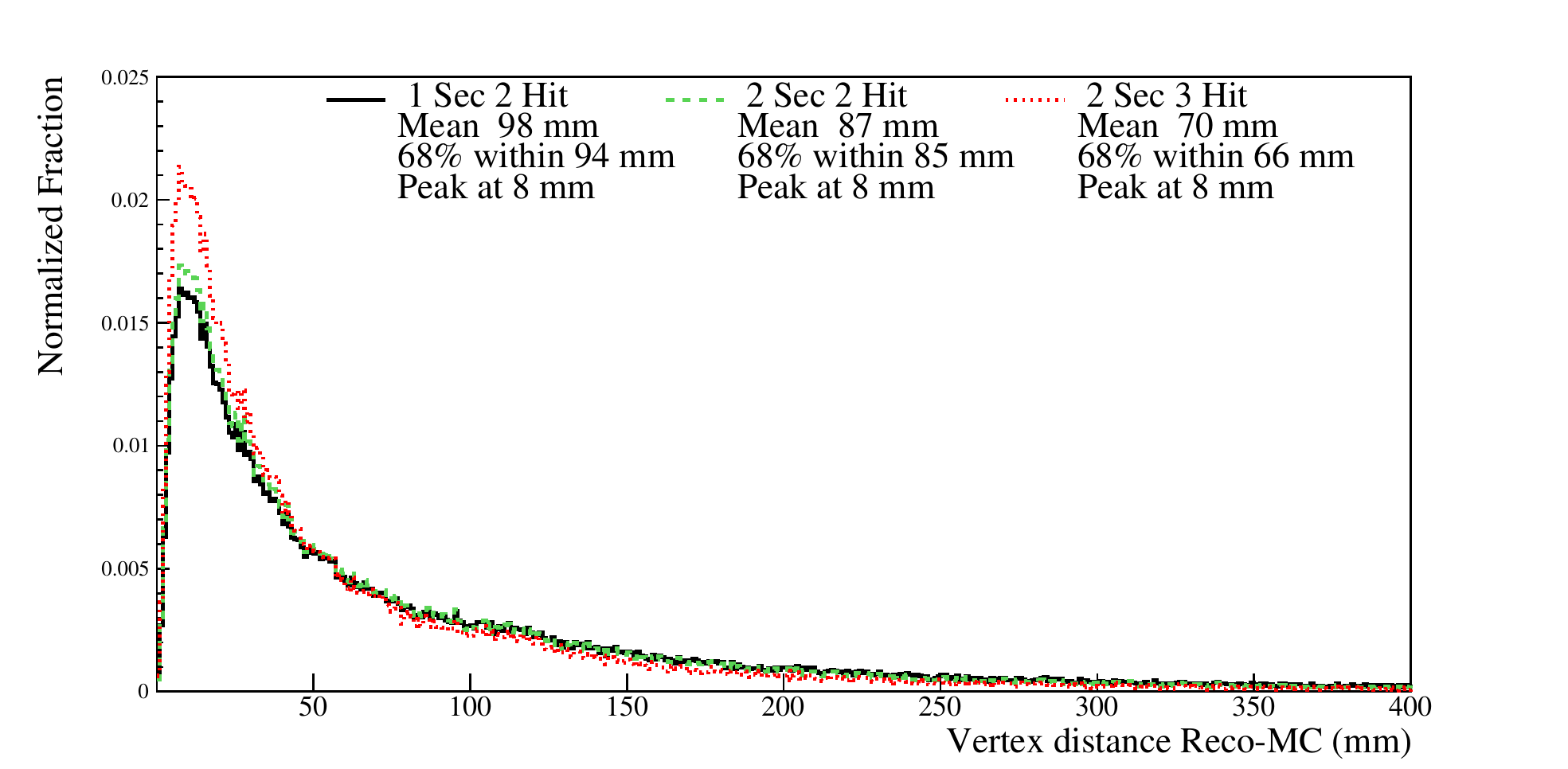}
\includegraphics[width=1\textwidth]{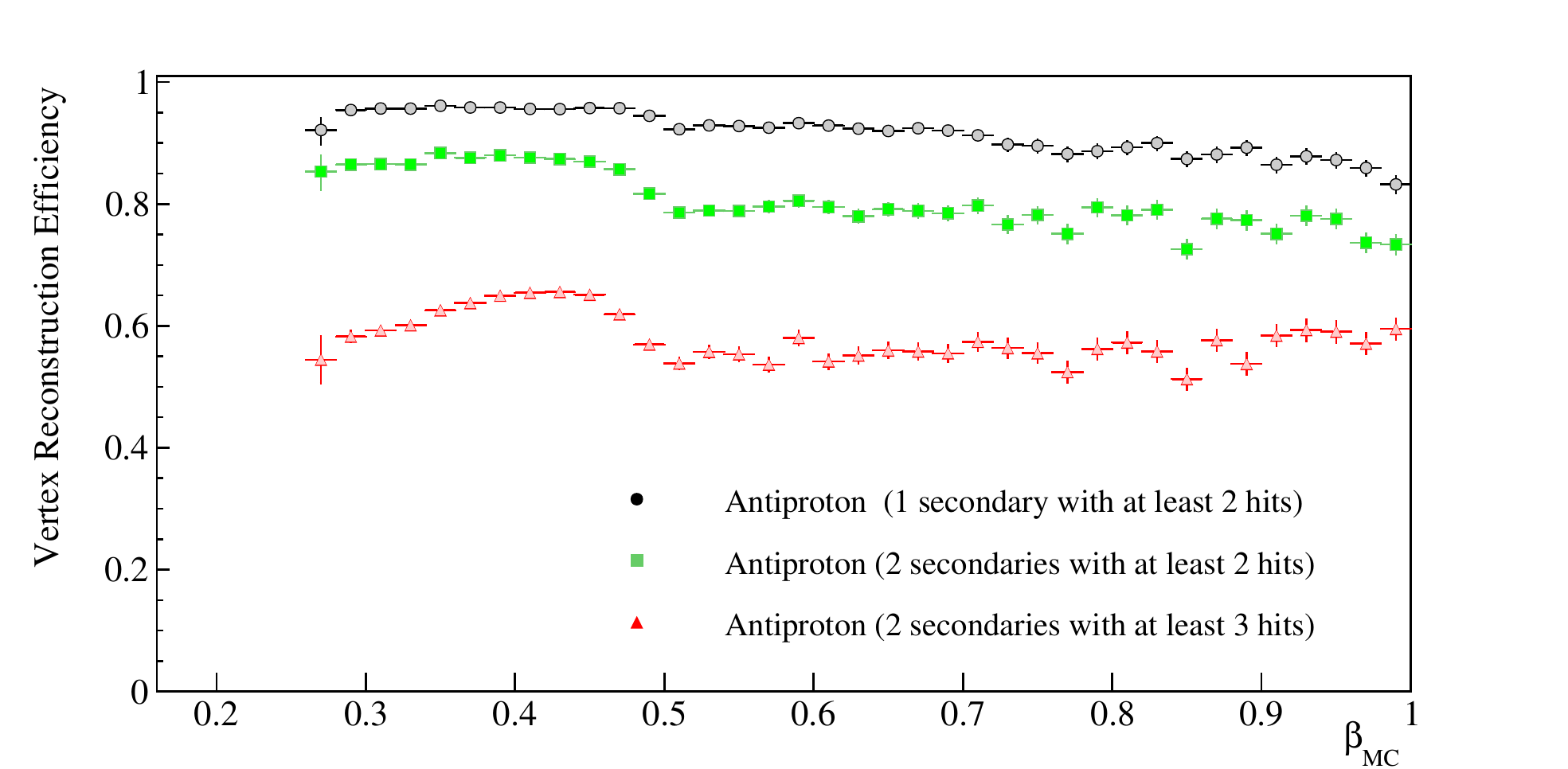}
\caption{
\textbf{Top panel:}  Antiproton vertex resolution performances with different requirements on the number of reconstructed secondary tracks and the number of hits associated with each track. \textbf{Bottom panel:}  Vertex reconstruction efficiency as a function of the Monte Carlo generated $\beta_{\text{MC}}$ for the star-finding algorithm and the different requirements on the number of reconstructed secondaries and hits. The initial sample was a set of antiprotons selected with the second trigger configuration
and the annihilation vertex inside the tracker volume according to Monte Carlo information. }
\label{fig12}
\end{figure}

In order to improve the vertex resolution, the position resolution 
was studied for different requirements on the number of reconstructed secondaries and hits for each secondary, and the results are shown in Figure ~\ref{fig12} (top panel).
The distribution for the events selected with the minimum requirements for a vertex reconstruction (solid line) has a $d_{68\%} = 94$\,mm that improves to $d_{68\%} = 85$\,mm requiring two secondary tracks with at least two hits each (dashed line) and to $d_{68\%} = 66$\,mm with two secondary tracks with at least three hits each (dotted line). However, using stricter requirements on the number of reconstructed secondary tracks decreases the number of selected events. This is demonstrated in the bottom panel of Figure ~\ref{fig12} 
for the case of antiprotons. The comparison sample is composed of events that pass the second trigger condition and meet the requirement that they annihilated inside the tracker volume according to Monte Carlo information. 
By requiring at least two reconstructed secondary tracks with at least two hits (squares), the total efficiency decreases by about $10\%$ with respect to the minimum quality cut. If events with at least two reconstructed secondaries and with at least three hits (triangles) were selected, the decrease of the total efficiency is almost $40\%$. These variations are not due to a decreasing performance of the reconstruction algorithm (the reconstruction efficiency is constant for this subset of events) but to the secondary multiplicity and 
the hit number on a secondary track

The width of the vertex resolution distribution is mainly a consequence of:
\begin{itemize}
    \item The spatial resolution of the digitized hit, which represents an intrinsic limitation to the resolution of the vertex position determination.  
     \item 
    The assumption in the reconstruction procedure that the particle trajectories are straight lines neglecting the effect of multiple scattering.
\end{itemize}

The tail of the vertex resolution distribution to high values is the result of various effects:
\begin{itemize}
    \item Wrong reconstruction of the primary trajectory: if the reconstructed primary trajeFctory significantly deviated from the correct one, the algorithm for the secondary search miscalculated the vertex since both the star finding and the Hough-3D algorithms uses the primary direction as a constraint.
    For example, a fraction of primary antinuclei (a few percent) hard-scattered on detector material nuclei, considerably changing their trajectory. The primary algorithm was not designed to account for this effect and, consequently, the reconstructed trajectory differs significantly from the true one for this small subset of events. 
    \item Double vertex events: it was found that a few percent of the events, which annihilate in the tracker volume have a double-vertex topology. For antiprotons, a double vertex can result from a secondary particle that undergoes a nuclear interaction inside the tracker producing a star of tertiary particles. For heavier antinuclei, in addition to this effect, also a two-step annihilation can take place. For example, the antiproton of an antideuteron annihilates first, producing a star, and the antineutron continues to travel and annihilates further down. This produces two different annihilation stars. 
    
    \item Vertices due to interactions of secondary particles: most of these events are antinuclei that annihilate in the outer ToF and produced secondaries which, in turn, interact inside the tracker, mimicking an annihilation star. In this case, the reconstructed primary track is the secondary particle created in the annihilation, which first hit the cube ToF. 
    For this type of event, the primary energy ratio $E_{\text{umb}}/E_{\text{cube}}$ is $\gg 1$ while for a slowing-down antinucleus that annihilates inside the tracker, the ratio is expected to be $\lesssim 1$. Thus, in order to reject this type of events, an energy ratio requirement of $E_{\text{umb}}/E_{\text{cube}}<2.5$ was introduced. This value rejects $95\%$ of vertices from secondary interactions while having a $98\%$ efficiency in selecting antinuclei annihilations. 
\end{itemize}

\begin{figure}
\centering
\includegraphics[width=1\textwidth]{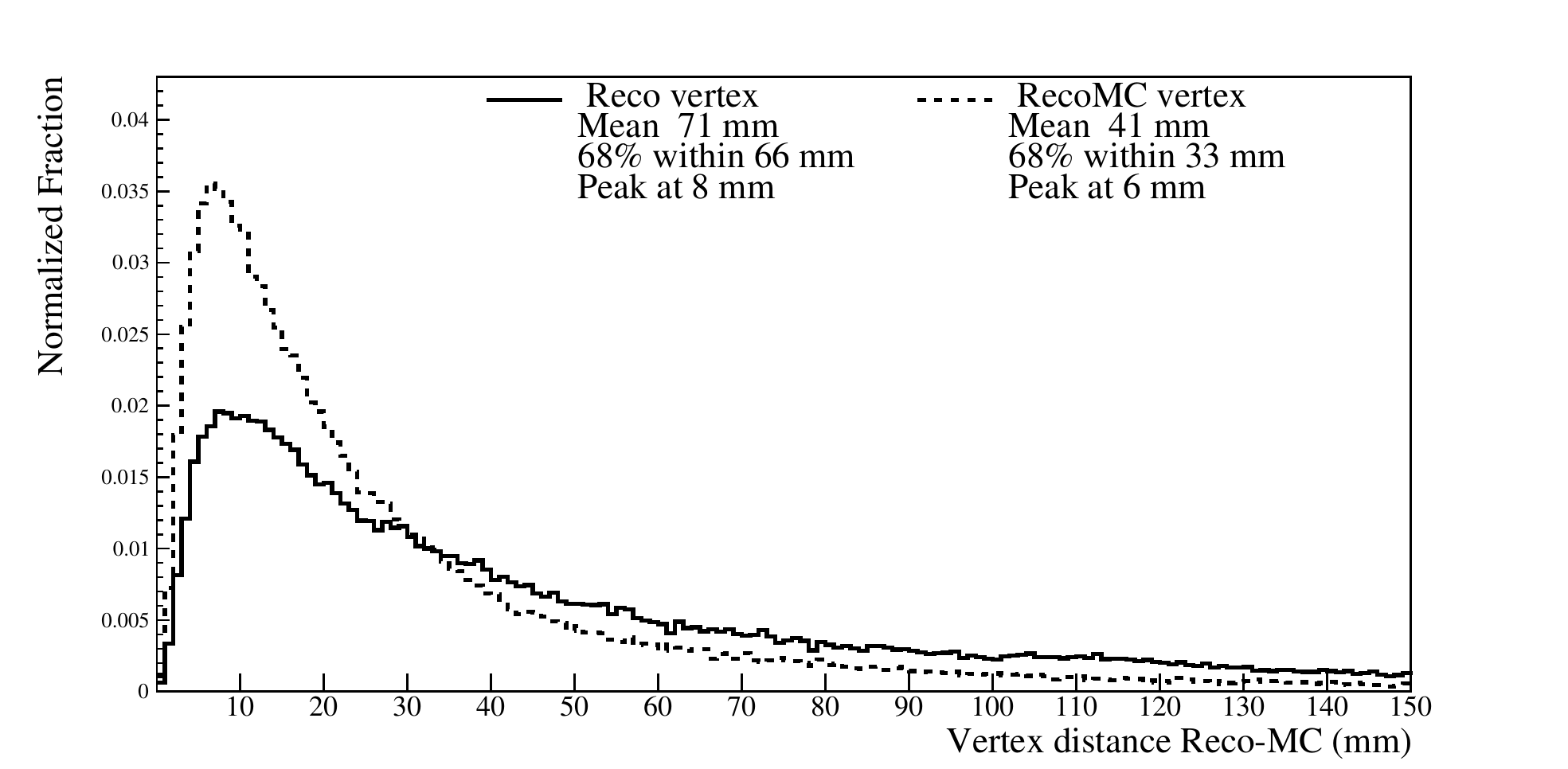}

\caption{Comparison between the vertex resolution performance obtained with the star-finding algorithm and its best possible performance. The dotted line refers to the vertex resolution that would be obtained if all the hits of the primary and of the secondary particles would be correctly associated with each track. }
\label{fig13}
\end{figure}

In order to quantify the performance of the vertex reconstruction algorithm, the best possible vertex resolution achievable with the star-finding algorithm was evaluated.
This best position reconstruction was obtained as follows: using the true Monte Carlo information, the digitized hits were associated with the corresponding primary and secondary tracks. Then, these hits were fit with straight lines, and the resulting trajectories were fed to the reconstruction algorithm to estimate the vertex position. 
The dotted distribution in Figure ~\ref{fig13} shows the resulting position resolution for a sample of antiprotons with the additional requirement of at least two secondaries with at least three hits each. The solid distribution shows the vertex resolution obtained applying the reconstruction 
with the same requirements on the number of secondary tracks and hits (dotted distribution in Figure ~\ref{fig12}).  
    Figure ~\ref{fig13} shows that the peaks of the two distributions are very similar, 8\,mm for the reconstruction and 6\,mm for the ideal case. As expected the overall performance for the ideal case is better than the one obtained with the reconstruction algorithm ($d^{\text{best}}_{68 \%} = 33$\,mm while $d^{\text{reco}}_{68\%} = 66$\,mm). Nonetheless, also in the ideal case, the distribution extends to high values indicating the intrinsic limitations. 
    
    \begin{figure}
\centering
\includegraphics[width=1\textwidth]{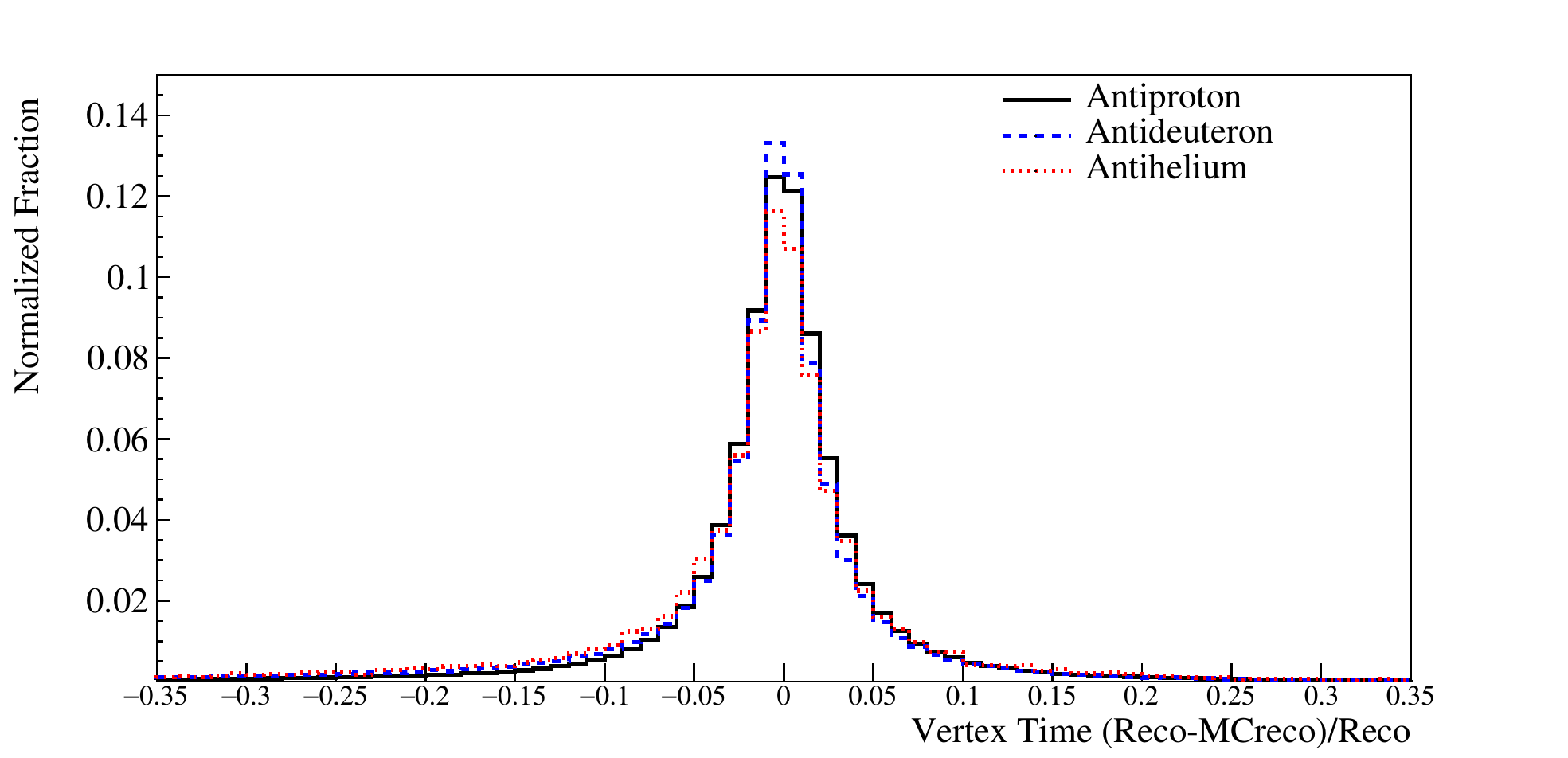}
\caption{Difference between the reconstructed and Monte Carlo annihilation time (normalized to the reconstructed time) for antiprotons (solid), antideuteron (dashed) and antihelium-3 (dotted).   }
\label{fig14}
\end{figure}

Figure ~\ref{fig14} shows the vertex time resolution, defined as the difference between the reconstructed and the Monte Carlo vertex time (normalized to the reconstructed time) for antiproton (solid line), antideuteron (dashed line), and antihelium-3 (dotted line). The distributions peak approximately at zero with an RMS of about $0.10$.

\subsection{Reconstruction Performance: First Flight}

For the first of the three planned GAPS flights, the tracker will be partially equipped with about 1000 Si(Li) detectors instead of 1440. In order to test the impact of this reduction on the reconstruction performance, a set of simulated antiprotons was produced, removing 100 modules (400 detectors) from the tracker. 
This configuration is equivalent to removing all modules of the two bottom tracker planes and all the side modules of the eighth plane.
This choice is based on the observation from antinuclei identification studies that the less-significant active volumes are those located in the bottom part and close to the lateral sides of the tracker. 
The removed active detectors will be substituted with non-active silicon material for the first flight.

\begin{figure}
\centering
\includegraphics[width=1\textwidth]{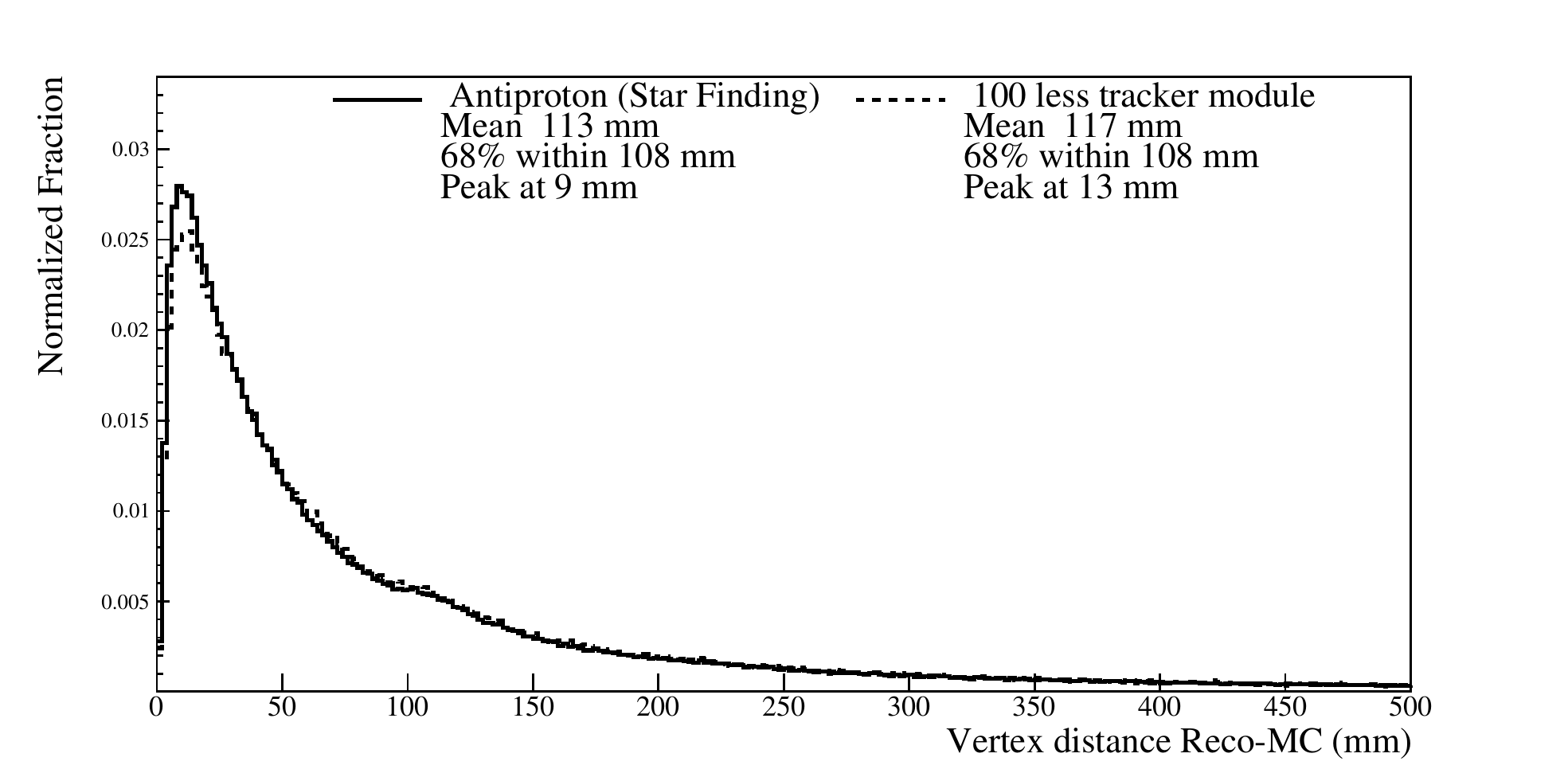}
\caption{Vertex resolution for different tracker configurations: with all the active modules (solid line) and with 100 inactive module (dashed line).
}
\label{fig16}
\end{figure}

It was found that the efficiency to reconstruct an annihilation vertex remains essentially the same when removing 100 modules as shown in Figure ~\ref{fig16}. The mean of the distribution for fewer modules is just a few\,mm higher and the peak increases from 8\,mm to 13\,mm, while the $d^{\text{reco}}_{68\%}$ value remains basically the same. 

\section{Conclusion}

The antinucleus annihilation reconstruction algorithm of the GAPS experiment was designed to clearly identify antinuclei annihilation patterns. The goal of the algorithm is to fully reconstruct the annihilation topology by estimating the position of the annihilation vertex and identifying the primary particle and secondary particle tracks. A custom algorithm was developed to search for star-like patterns. A modified Hough-3D transform algorithm was also developed and tested, but it was found to have inferior performance as compared to the star-finding algorithm.

The performance of the antinucleus annihilation reconstruction was studied using the GAPS simulation software based on the {\tt Geant4} toolkit. 
The star-finding algorithm was able to reconstruct the correct vertex position with an accuracy of 11\,cm (improvable to 7\,cm with more stringent but less efficient requirements on the topology) for more than 68\% of the reconstructed annihilations with $\approx 90\%$ efficiency in identifying antinuclei annihilating in the apparatus. This topology information will be one important ingredient for the particle identification of the GAPS experiment. For example,  
the average difference between the range of antideuterons and antiprotons with nearly vertical incidence and comparable velocity $\beta < 0.4$ is greater than 12 cm. 

The performance of the reconstruction algorithm was also studied for the case of a reduced number of tracker modules, as is expected to be the case for the first GAPS flight. Also for this configuration 
very good performance was obtained: a similar vertex reconstruction efficiency with a slight decrease in the vertex position resolution of about a few millimeters. 

\section{Acknowledgment}

This work is supported in the U.S. by NASA APRA grants (NNX17AB44G, NNX17AB45G,279NNX17AB46G, and NNX17AB47G) and in Japan by JAXA/ISAS Small Science Program FY2017.280P. von Doetinchem received support from the National Science Foundation under award PHY-2811551980. F. Rogers is supported through the National Science Foundation Graduate Research Fellowship under Grant No. 1122374. This work is supported in Italy by Istituto Nazionale di Fisica Nucleare (INFN) and by the Italian Space Agency through the ASI INFN agreement n. 2018-28-HH.0: “Partecipazione italiana al GAPS - General AntiParticle Spectrometer”. The technical support and advanced computing resources from the University of Hawaii Information Technology Services – Cyberinfrastructure are gratefully acknowledged.

\appendix

\section{Estimation of Kinematic Parameters of Tracks}
\label{AppendixA}
The first step in the estimation of the kinematic parameters of the primary and secondary tracks is the determination of the particle trajectories. 
These trajectories are assumed to be straight lines, i.e.:
\[
    \mathbf{p} = \mathbf{d} \cdot l + \mathbf{a} 
\]
with $\mathbf{p}$ a point in three-dimensional space, $\mathbf{d}$ a vector that represents the line direction, $\mathbf{a}$ the anchor point, and $l$ a parameter that represents the path length along the line, relative to the anchor point. 
Two coordinates ($a_x$,$a_y$) define the anchor point position in the reference plane, while two polar angles ($\theta$, $\phi$), relative to the normal plane, define the line direction.  
Each straight line is uniquely described by a four-element state vector $\mathbf{q}=\{ a_x, a_y, \theta, \phi \}$. 

The trajectory state vector is determined by a least-squares minimization procedure, where the minimized quantity is defined as:
\[
\chi^2 = \sum_{i=1}^n \left[ \left( \frac{\Delta x'_i(\mathbf{q})}{\delta_{x,i}} \right)^2 + \left( \frac{\Delta y'_i(\mathbf{q})}{\delta_{y,i}}\right)^2 \right]
\]
with $n \geq 2$ the number of hits classified as belonging to the same track, 
$\Delta x'$ and $\Delta y'$ the distances between the evaluated impact positions on the detection plane and the measured hit positions, 
$\delta_{x,i}$ and $\delta_{y,i}$ the spatial residuals are accounting for the physical dimensions of the detecting elements.

The impact position on a detector volume is the intersection point between the line approximating the particle trajectory and a flat surface co-planar with the detector, centered at its midpoint. This flat surface determines the reference system ($x', y'$) used to represent the intersection point.

After the track fitting, requirements on the track quality of the track candidate are applied: 

\begin{enumerate}

\item The $\chi^2$ of the reconstructed track has to be smaller than $\chi^2_{\text{cut}}$. This selection removes tracks resulting from particles that suffered significant multiple scattering. The value of $\chi^2_{\text{cut}}$ was set to 6, providing a Monte Carlo estimated efficiency of $98\%$;

\item The track has to have a length\footnote{The track length is the spatial distance between the positions of the first and last hits. } greater than $l_{\text{cut}}$ and has to traverse at least two detection planes. The value of $l_{\text{cut}}$ was set to $10$\,cm. These cuts removed secondary track candidates that only had two adjacent ToF hits.

\end{enumerate}
If these requirements are not satisfied, the track candidate is rejected.

The covariance matrix of the state vector fit is used to propagate the errors when calculating track-related quantities.
Once the trajectory is determined, kinematic parameters associated with the track, like the path length $l_i$ and the (d$E/$d$x)_i$ values associated to the $i^{\text{th}}$ measured hit belonging to the track, are evaluated.  
If the track includes at least two time ($t$) measurements, the track velocity is evaluated 
via the linear relation $t_i = l_i/\beta c +t_0$, 
with $\beta = v /c $ ($v$ particle velocity and $c$ speed of light). If more than two time measurements are associated to the track, a least square minimization is used.

\section{Hough-3D Transform}
\label{Hough3DSec}

As a global method to the hit classification problem, an algorithm based on the Hough transform \cite{HoughOriginal},  adapted from \cite{ipol.2017.208}, was developed.
This global algorithm searches in the parameter
space the set of parameters $(\mathbf{a}, \mathbf{d})$ (\ref{AppendixA}) which best describes a group of aligned points in the 3D space.
The most relevant parameter of the algorithm is the segmentation of the parameter space, which depends substantially on the spatial resolution. 

The original code \cite{ipol.2017.208} was modified to account for the specific pattern of the annihilation process in the GAPS experiment. It is required that the track candidates have to originate not more than a few centimeters from the primary track.  This constraint reduces drastically the selection of hits randomly aligned on a straight track.  
The performance of the algorithm is mostly limited by the misalignment of hits caused by multiple scattering process. This problem is reduced by using a sufficiently large segmentation of the parameter space.
In this work, the segmentation was
chosen to be 160\,mm, i.e., approximately the spatial resolution of the ToF hit.

The Hough transform was applied recursively:  when a set of hits is associated with a track candidate, the hits are removed, and the algorithm applied again.  
Finally, a  straight-line fit is performed on each hit set in order to estimate the secondary trajectories.

It was found that this Hough transform algorithm is quite sensitive to 
the small number of hits generated by secondary tracks in the GAPS apparatus and to the different spatial resolutions of the tracker and the ToF hits. The required increase of the space segmentation had to be balanced with the increased probability of wrongly associating hits to secondary tracks. This limited the performances of the Hough-3D transform with respect to the custom algorithm as shown in Secs.~\ref{secefficiecy} and \ref{SecVertexRes}.

\bibliography{mybibfile}

\end{document}